\documentclass[epj]{svjour}
\newcommand{\be}{\begin{eqnarray}}
\newcommand{\ee}{\end{eqnarray}}

\newcommand{\ave}[1]{\left\langle #1 \right\rangle}
\usepackage{graphics}
\usepackage{epsfig}
\begin{document}
\title{Resonances and fluctuations at SPS and RHIC}
\author{Giorgio Torrieri}
%
\offprints{}          
\institute{McGill University}
\date{Received: 29 July 2006 / Revised version: date}
%
\abstract{
We perform an analysis of preliminary data on hadron yields and fluctuations  within the Statistical hadronization ansatz.
We describe the theoretical disagreements between different statistical models currently on the market, and show how the simultaneous analysis of yields and fluctuations can be used to determine if one of them can be connected to underlying physics.   We perform such an analysis on preliminary RHIC and SPS A-A data that includes particle yields, ratios and event by event fluctuations.
We show that the equilibrium statistical model can not describe the $K/\pi$ fluctuation measured at RHIC and SPS, unless an unrealistically small volume is assumed.  Such small volume then makes it impossible to describe the total particle multiplicity.
The non-equilibrium model,on the other hand, describes both the $K/\pi$ fluctuation and yields acceptably due to the extra boost to the $\pi$ fluctuation provided by the high pion chemical potential.        We show, however, that both models significantly over-estimate the $p/\pi$ fluctuation measured at the SPS, and speculate for the reason behind this.  
\PACS{25.75.-q,24.60.-k,24.10.Pa
     } 
} 
\maketitle
\section{Introduction}
One of the main objectives of heavy ion physics is to study the collective properties of QCD matter.    It's equation of state, transport coefficients and phase structure, and the dependence of these on energy and system size.

Thus, the natural approach to study soft particle production in heavy ion collisions is through statistical mechanics techniques.   
Such an approach has a long and illustrious history \cite{Fer50,Pom51,Lan53,Hag65}.      A consensus has developed that the statistical hadronization model can indeed fit most or all particles for AGS,SPS and RHIC energies
\cite{jansbook,bdm,cleymans,barannikova,gammaq_energy,gammaq_size,becattini}.

The statistical model obtains particle yields by assuming entropy to be maximized given the constraints imposed by energy and quantum number
conservation.  These constraints can either be imposed rigorously, as required for closed equilibrated systems, or on average, as required for a sub-system equilibrated with an unobserved ``bath''.   Full energy and quantum number conservation is usually referred to as the micro-canonical ensemble, while
the Canonical (C) and Grand-canonical (GC) ensembles assume that, respectively, energy and other conserved quantities can vary via system-bath exchange.
In this work, we shall concentrate on the GC ensemble, as we see it as most appropriate for describing the statistically hadronizing
fireballs produced in heavy ion collisions.  

Our approach is not universally agreed on by the heavy ion community;
In fact, noteworthy attempts were made to explain the dependence of certain observables w.r.t. energy and system size 
(The ``horn'' structure in the $K^+/\pi^+$ ratio dependence on energy \cite{horn}, as well as the strange particle enhancement at larger energies and system sizes \cite{stenhsps,stenhrhic}) on the limits reached in different regimes;  Averages of observables over {\em all} events are the same in all three ensembles in the limit
of either a ``large'' system (high energy and system volume) or an observed small sub-system of a large system (mid-rapidity).
{\em away} from these limits (low energy, p-p and p-A collisions), the three ensembles give very different results, and, given the low multiplicity of strange
particles in these collisions, it is necessary to enforce exact conservation to correctly count the particle states (Canonical or micro-canonical ensemble).
The onset of strangeness enhancement \cite{strangeness_canonical}, and perhaps structures such as the horn, have been speculated to arise from the transition between these limits.

Both the horn \cite{horn_theory} and strangeness enhancement \cite{jansbook} are also regarded as indicators of a deconfined state of matter.   If this is correct, than the energy and system size where they develop indicates the point at which the system can be described as an equilibrated deconfined Quark-Gluon Plasma (QGP).

Distinguishing between these two explanations is a {\em phenomenological} question: The ensemble assumption is a falsifiable one, in the sense that a physically inappropriate ensemble is very unlikely to describe the available experimental data, if the data includes yields and event-by-event fluctuations
\cite{nogc1,nogc2,nogchg}.   Hence, a statistical model analysis at  different energies and system sizes, of both yields and event-by-event fluctuation observables, will isolate the statistical physics (if any) responsible for the observed features.   The present work is a step in this direction, focusing on a Grand Canonical analysis of top energy RHIC ($\sqrt{s}=200$ GeV/A Au-Au collisions) and SPS ($\sqrt{s}=17.6$ GeV/A Pb-Pb collisions) heavy ion data.

Yields and fluctuations in the GC ensemble are calculated via the quantum statistic ``ideal gas'' formulae
\be
\label{yield_formula}
\ave{ N_i}
  &=&  \int { g_i V\over (2\pi)^3}\, {4 \pi p^2 d p \over  \lambda_i^{-1} e^{\sqrt{p^2+m_{i}^{2}}/T}\pm 1}\\\ave{(\Delta N_i)^2} & = & \lambda_i{\partial \over
                 \partial \lambda_i} \ave{ N_i}
\ee
Where $g_i$ is the degeneracy, $V$ the system volume, the upper sign is for Fermions and the lower sign for bosons.  It is sometimes confusing that states in a ``strongly interacting'' system can be
accurately predicted via the ideal gas ansatz.  The key insight \cite{Hag65} is that accounting for {\em all} strong excitations (resonances) is equivalent to counting the QCD ``energy levels''.    Since hadronic interactions are resonance dominated (a consequence of the
confining nature of QCD) this is a good approximation.   Lattice studies have confirmed the validity of this approach quantitatively \cite{redlichlat}.

The final state yield of particle $i$ is computed by adding the direct
yield and all resonance decay feed-downs.
\be
 \label{resoyield}
\ave{N_i}_{\rm tot} & = &
\ave{ N_i} + \sum_{{\rm all}\;j \rightarrow i}
B_{j \rightarrow i}  \ave{N_j} \\
 \ave{(\Delta N_{ i})^2}_{\rm tot} 
& = & \ave{(\Delta N_{ i})^2}+  \sum_{{\rm all}\;j \rightarrow i} B_{j\to i}^2 \ave{(\Delta N_j)^2} \\ &+&
 \sum_{{\rm all}\;j \rightarrow i} B_{j\to i}(1-B_{j\to i})\ave{N_j} \nonumber
\ee
The parameter $\lambda_i$ corresponds to the particle fugacity, related to the chemical potential by $\lambda_i=e^{\mu_i/T}$.  Provided the law of mass action holds, it should be given by the product of charge fugacities (flavor, isospin etc.).  It can be parametrized it in terms of equilibrium fugacities $\lambda_i^{\mathrm{eq}}$ and phase space occupancies $\gamma_i$.   A hadron $i$ with $q (\overline{q})$ light quarks, $s (\overline{s})$ strange quarks and isospin $I_3$ has fugacity 
\begin{eqnarray}
\label{chemneq}
\lambda_i = \lambda_i^{\mathrm{eq}}
\gamma_q^{q+\overline{q}} \gamma_s^{s+\overline{s}}
\phantom{A},\phantom{A}
\lambda_i^{\mathrm{eq}}=\lambda_{q}^{q-\overline{q}} \lambda_{s}^{s -
\overline{s}}
\lambda_{I_3}^{I_3} 
\end{eqnarray}
The temperature and chemical potentials can be obtained from data by doing a $\chi^2$ fit \cite{pdg}
(The chemical potentials for strangeness, $\lambda_s$, and isospin, $\lambda_{I3}$, are usually obtained by requiring strangeness and $Q/B$ to be conserved).

If the system is in chemical equilibrium then detailed balance requires that $\gamma_q=\gamma_s=1$.    Assuming $\gamma_q=1$ and fitting
ratios gives the $T \sim 160-170$ MeV freeze-out temperature typical of chemical equilibrium freeze-out models at SPS and RHIC \cite{bdm,cleymans,barannikova}.  
Some fits \cite{cleymans,becattini} also allow for a kinetically out of equilibrium 
 strangeness quantum number, which is found to be at chemical equilibrium ($\gamma_s=1$) at RHIC and slightly below equilibrium ($\gamma_s<1$) at SPS.

In a system expanding and undergoing a phase transition, however, the condition of chemical equilibrium 
no longer automatically holds, so one has to allow for the possibility that both $\gamma_s$ {\em and} $\gamma_q \ne 1$.
In particular, if the expanding system undergoes a fast phase transition from a QGP to a hadron gas, chemical non-equilibrium  \cite{jansbook} and super-cooling \cite{csorgo} can arise due to entropy conservation:
By dropping the hadronization temperature to $\sim 140$ MeV and oversaturating the hadronic phase space above equilibrium ($\gamma_q \sim 1.5,\gamma_s \sim 2$), it is possible to match the entropy of a hadron gas with that of a system of nearly massless partons \cite{jansbook}.
These are exactly the values found for $\gamma_q$ and $T$ at SPS and RHIC in fits where $\gamma_q$ was a fit parameter \cite{gammaq_energy,gammaq_size}.

The ``$\gamma_q=1$'' and ``$\gamma_q$ fitted'' approaches are different {\em models} of how hadrons are produced in heavy ion collisions.  They are both based
on statistical mechanics, yet differ in the physics underlying it.
They are, in principle, distinguishable experimentally but have not been to date.

Fig. 1  in \cite{gammaq_energy} illustrates why an experimental test of either of these models is non-trivial:
An increase in temperature acts in a very similar way on particles and anti-particles as an
increase in $\gamma$: The abundances of both go up.
The amount in which relative particle abundances go up is different, making the two models distinguishable by yields alone.  However, this difference is not enough to convincingly disentangle the two models, given the large experimental errors in measurements of yields and ratios.

To falsify one of the models it is necessary to consider event-by-event fluctuations as well
as average particle yields \cite{prcfluct}.
If $\gamma_q$ becomes as large as claimed in \cite{gammaq_energy}, the pion chemical potential $\lambda_{\pi}$
approaches $e^{m_\pi/T}$.   Near this limit (corresponding to the critical density for pion B-E condensation) it can be shown \cite{prcfluct} that $\ave{N_\pi}$ converges but $\ave{(\Delta N_\pi)^2}$ diverges as
\begin{equation}
\label{divergence}
  \ave{(\Delta N)^2} \sim \epsilon^{-1/2}.
\end{equation}
where $\epsilon = 1 - \lambda_\pi e^{-m_\pi/T}$. 
thus, while $T$ and $\gamma_q$ are correlated in yields, they are anti-correlated in fluctuations.
The measurement of a yield and fluctuation constrains both to a high precision.
At least one of the models will be ruled out by inclusion of a fluctuation in the fit.
Inclusion of more than one fluctuation can be used to test the remaining model.

The question of chemical equilibration is is closely related to the accounting for hadronization when modeling freeze-out.
While a quantitative treatment of this topic is still lacking, it is obvious that if the freeze-out temperature is higher (i.e., $\gamma_q=1$), the effect of hadronic interactions between chemical and thermal freeze-out is greater than if the system freezes out from a super-cooled state ($\gamma_q>1$).

It is also unclear what, if any, is the effect on observables of the evolution between hadronization (the formation of hadrons as effective degrees of freedom) and freeze-out (the moment when all hadrons decouple) \cite{hiranogyul}.
As shown in \cite{hydroheinz}, the discrepancy with experiment would lessen if chemical and kinetic freeze-out were to coincide.
While, as claimed in \cite{hydroheinz}, such a high freeze-out temperature would spoil agreement with particle spectra, fits based on a single freeze-out model prescription have shown 
that, provided a correct treatment of resonances is maintained, particle spectra are compatible with simultaneous freeze-out
\cite{florkowski,us_sps,us_rhic}.   

A way to distinguish between these scenarios is to directly measure the abundance of resonances.  Here, the situation becomes even more ambiguous:
As pointed out in \cite{usresonances}, resonance abundance generally depends on two quantities:  $m/T$, where $m$ is resonance mass and the chemical freeze-out temperature, as well as $\tau \Gamma$, where $\Gamma$ is the resonance width and $\tau$ is the reinteraction time.  Observing two ratios where the two particles have the same chemical composition, but different $m$ and $\Gamma$, such as $\Lambda(1520)/\Lambda$ vs $K^*/K$, or $\Sigma(1385)/\Lambda$ vs $K^*/K$, could therefore be used to extract the magnitude of the freeze-out temperature and the re-interaction time.
Studies of this type are still in progress;    
As we will show, $\Lambda(1520)/\Lambda$ and $K^*/K$ seem to be compatible with sudden freeze-out, provided freeze-out happens in a super-cooled over-saturated state \cite{jansbook,gammaq_energy,gammaq_size}.
Other preliminary results, such as $\rho/\pi$, $\Delta/p$ \cite{barannikova}, and now $\Sigma^*/\Lambda$ \cite{salur} seem to be produced \textit{in excess} of the statistical model \cite{barannikova,fachini,salur}, both equilibrium and not.     It is difficult to see how a long re-interacting phase would produce such a result:  Resonances whose interaction cross-section is small w.r.t. the timescale of collective expansion would generally be depleted by the dominance
of rescattering over regeneration processes at the detectable (on-shell) mass range. More strongly interacting resonances would be re-thermalized at
a smaller, close to thermal freeze-out temperature.   Both of these scenarios would generally result in a suppression, rather than an enhancement, of directly detectable resonances.    Transport model studies done on resonances generally confirm this \cite{urqmd1,urqmd2}.
Yet the only resonance, so far, found to be strongly suppressed w.r.t. expectations is the SPS $\Lambda(1520)$ \cite{resosps}, and even that measurement is to date preliminary.

The observation of an enhanced $\mu^+ \mu^-$ continuum around the $\rho$ peak \cite{na60} has been pointed to as evidence of $\rho$ broadening, which
in turn would signify a long hadronic re-interaction phase \cite{na60}.
The absence of a broadening in the {\em nominal} peak itself, prevents us from considering this as
the unique interpretation of experimental data.  Moreover, even a conclusive link of broadening with hadronic re-interactions would still give no indication to the length of the hadronic rescattering period.   Nor it would resolve the discrepancies pertaining hadronic resonances encountered in the previous paragraph;
The lack of modification in either mass or width, between p-p and Au-Au seen so far \cite{fachini,salur} is only compatible with the NA60 result
 provided the
$\rho$ is {\em very} quick to thermalize, so \cite{fachini} sees only the $\rho$ s formed close to thermal freeze-out.
  But, as argued in the previous paragraph, that under-estimates the abundance of the $\rho$ and other resonances, since the $\rho/\pi,K^*/K$, and even $\Lambda(1520)/\Lambda$ ratios point to a freeze-out temperature significantly above the 100 MeV, commonly assumed to be the ``thermal freeze-out'' temperature in a staged freeze-out scenario.

To help resolve this ambiguity, we aim to {\em directly infer} the magnitude of hadronic reinteraction by combining, within the same analysis,
an observable sensitive to the chemical freeze-out resonance abundance with the direct observation of resonances.
As shown in \cite{jeonkochratios}, the measurement of fluctuations of a ratio is such an observable.

The fluctuation of a ratio $N_1/N_2$
can be computed from the fluctuation of the denominator and the numerator \cite{jeonkochratios}
($\sigma_{X}^2=\ave{(\Delta X)^2}/\ave{X}$):
\begin{equation}
\label{fluctratio}
\sigma_{N_1/N_2}^2
= \frac{\ave{(\Delta N_1)^2}}{\ave{N_1}^2}
+ \frac{\ave{(\Delta N_2)^2}}{\ave{N_2}^2}
- 2 \frac{\ave{\Delta N_1 \Delta N_2}}{\ave{N_1}\ave{ N_2}}.
\end{equation}
Note the appearance of a negative correlation term  between $N_1$ and $N_2$ stemming from a common
resonance feed-down ($\Delta \rightarrow p \pi$ will be a source of correlation between $N_{p}$ and $N_{\pi}$).
\begin{equation}
\ave{\Delta N_1 \Delta N_2} \approx \sum_j b_{j \rightarrow 1 2} \ave{N_j}
\end{equation}
This correlation term is an invaluable phenomenological resource, since it is sensitive to resonance abundance at {\em chemical} freeze-out \cite{jeonkochratios};  Further reinteraction, provided resonances are not kicked out of the detector's acceptance region \cite{prcfluct}, preserve the correlation even when the original resonance
stops being reconstructible.   Thus, a measurement of the ratio fluctuation together with the directly detectable resonance yield gauges the amount
of re-interaction between the thermal and chemical freeze-out.   If the reinteraction period is short, both observables should be described by the same statistical model parameters

The main experimental problem with fluctuation measurements is the vulnerability to effects resulting from limited detector acceptance.
This difficulty can be lessened, to some extent, by considering ``dynamical'' fluctuations, obtained by subtracting a ``static'' contribution which
should be purely Poisson in an ideal detector.
\begin{equation}
\sigma^{dyn}=\sqrt{\sigma^2-\sigma_{stat}^2}
\end{equation}
$\sigma_{stat}$, usually obtained through a Mixed event approach \cite{pruneau},
includes a baseline Poisson component, which for a ratio $N_1/N_2$ can be modeled as
\begin{equation}
\label{poissrat}
\sigma_{stat}^2 = \frac{1}{\ave{N_1}}+\frac{1}{\ave{N_2}}
\end{equation}
 as well as a contribution from detector efficiency and kinematic cuts.    Provided certain assumptions for the detector response function hold (see appendix A of \cite{pruneau}), subtracting
$\sigma_{stat}$ from $\sigma$ should yield a ``robust'' detector-independent observable.

More complicated to deal with are detector acceptance effects affecting particle {\em correlations} (the probability for both resonance decay products to be within the detector acceptance region) \cite{prcfluct}.   These were not corrected for in the present study;   RHIC data-points do not include this term, while at SPS, due to it's large acceptance, this term is likely to be less important.

Related to the time-scale of the thermal freeze-out is the question of the total normalization of the system.
This quantity can be obtain in fits by fitting yields of particles rather than ratios.
It is physically important because it is connected to the system's volume at chemical freeze-out, which is in turn important to gauge the relative importance of the hadronic phase in the system's dynamics \cite{hiranogyul}.  It is also necessary to obtain the thermal energy and entropy content of the system
\cite{gammaq_energy}, and it's scaling with centrality.

On the other hand, normalization introduces an additional source of inter-parameter correlation, since it scales in the same way as $T$ and $\gamma_q$.
Measuring fluctuations of ratios in addition to yields and rations, once again, can be used to disentangle these three quantities. Note, from Eqs. \ref{fluctratio} and \ref{poissrat} that fluctuations of ratios, while independent of the {\em volume fluctuation} scale as the inverse of the absolute normalization, $\sigma_{N_1/N_2} \sim (\ave{V}T^3)^{-1}$.

Hence an increase in volume makes the yields go up, does not affect the {\em average} ratios, but decreases the {\em event-by-event fluctuation of ratios}.   An increase in temperature generally affects both yields and ratios, increases the yields and decreases the fluctuations.   An increase in $\gamma_q$ increases yields, baryon to meson ratios {\em and} fluctuations.   It is therefore not difficult to fix all three parameters with a data-sample of relatively few data-points, provided both yields, ratios and fluctuations are present.

\section{Fit to RHIC and SPS data}
\begin{table*}
\caption{Best fit parameters at SPS Pb-Pb  $\sqrt{s}$=17.3 GeV and RHIC Au-Au $\sqrt{s}$=200 GeV  collisions}
\label{restable}       
\begin{center}
\begin{tabular}{|l|l|l|l|l|}
\hline\noalign{\smallskip}
& \multicolumn{2}{|c|}{  RHIC Au-Au $\sqrt{s}$=200 GeV  } &  \multicolumn{2}{|c|}{  SPS Pb-Pb $\sqrt{s}$=17.3 GeV  }\\ \hline
 & $\gamma_q \ne 1$  &  $\gamma_q = 1$   &  $\gamma_q \ne 1$  & $\gamma_q = 1$   \\  \noalign{\smallskip}\hline
T [MeV] & 143.3 $\pm$ 2.5 & 159.6 $\pm$ 6.9 & 138.2 $\pm$ 2.3 & 151.3 $\pm$ 3.8  \\
$\mu_B=3 T \ln(\lambda_q)$ [MeV] & 20.8 $\pm$ 3.9 & 23.2 $\pm$ 3.4 & 231.2 $\pm$ 19.1 & 248.0 $\pm$ 25.1 \\
$\mu_s=T \ln(\lambda_q \lambda_s^{-1}$ [MeV] & 4.5  $\pm$ 1.1 & 5.1 $\pm$ 0.8 & 58.6 $\pm$ 0.78 & 60.5  $\pm$ 9.3 \\
$\mu_{I3} = T \ln \lambda_{I3}$ [MeV] & -0.4 $\pm$ 0.1 & -0.7 $\pm$ 0.2 & -4.1 $\pm$ 1.1 & -8.8 $\pm$ 2.1\\
$\gamma_q$ & 1.541 $\pm$ 0.001 & 1 & 1.645 $\pm$ 0.003 & 1. \\
 $\gamma_s$ & 1.980 $\pm$ 0.158 & 1.083 $\pm$ 0.112 & 1.577 $\pm$ 0.116 & 0.815 $\pm$ 0.067 \\
Normalization ($fm^3$) & 1356 $\pm$ 395 & 1689 $\pm$ 488 & 2540 $\pm$ 658 & 3746 $\pm$ 444\\
\hline
\end{tabular}
\end{center}
\end{table*}

We have performed fits for both SPS and RHIC energies, using publicly available statistical hadronization software \cite{share,sharev2}.  For RHIC, we have used the same data sample as in \cite{sqm2006}, including $\sigma^{dyn}_{K^+/\pi^+}$ and  $\sigma^{dyn}_{K^-/\pi^-}$ \cite{supriya}.   For SPS yields, we used the same data sample as in \cite{gammaq_energy}, augmented by preliminary fluctuation measurements of $\sigma^{dyn}_{(K^++K^-)/(\pi^++\pi^-)}$ and $\sigma^{dyn}_{(p+\overline{p})/(\pi^++\pi^-)}$ \cite{spsfluct}.   It should be noted that the model under consideration is far more plausible when applied to RHIC than to SPS data, since the fluctuation measurement in \cite{spsfluct} encompasses a large acceptance region with non-trivial momentum cuts, that only with the greatest hesitation can be imagined as a sub-system in equilibrium with a bath.   

Our fit parameters include the normalization (hopefully related to the system ``volume'' at chemical freeze-out), temperature, $\lambda_{q,s,I_3}$ and $\gamma_{q,s}$.
We also require, by implementing them as data-points, strangeness ($\ave{s-\overline{s}}=0 \pm 0.01$), charge and baryon number ($\ave{Q}/\ave{B}=Z/A \pm 0.01$) conservation.

\begin{figure*}
\epsfig{width=8.cm,clip=,figure=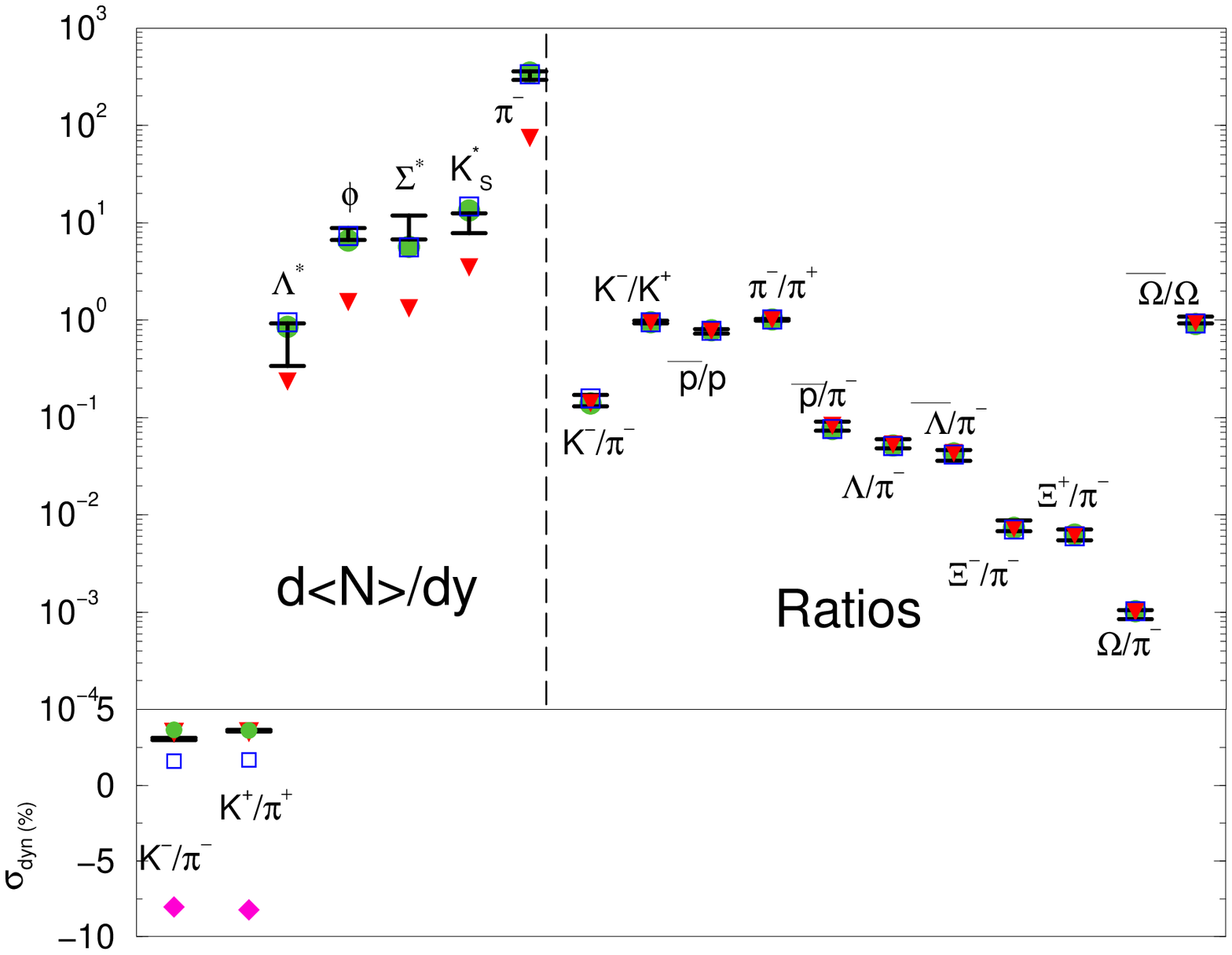}
\epsfig{width=8.1cm,clip=,figure=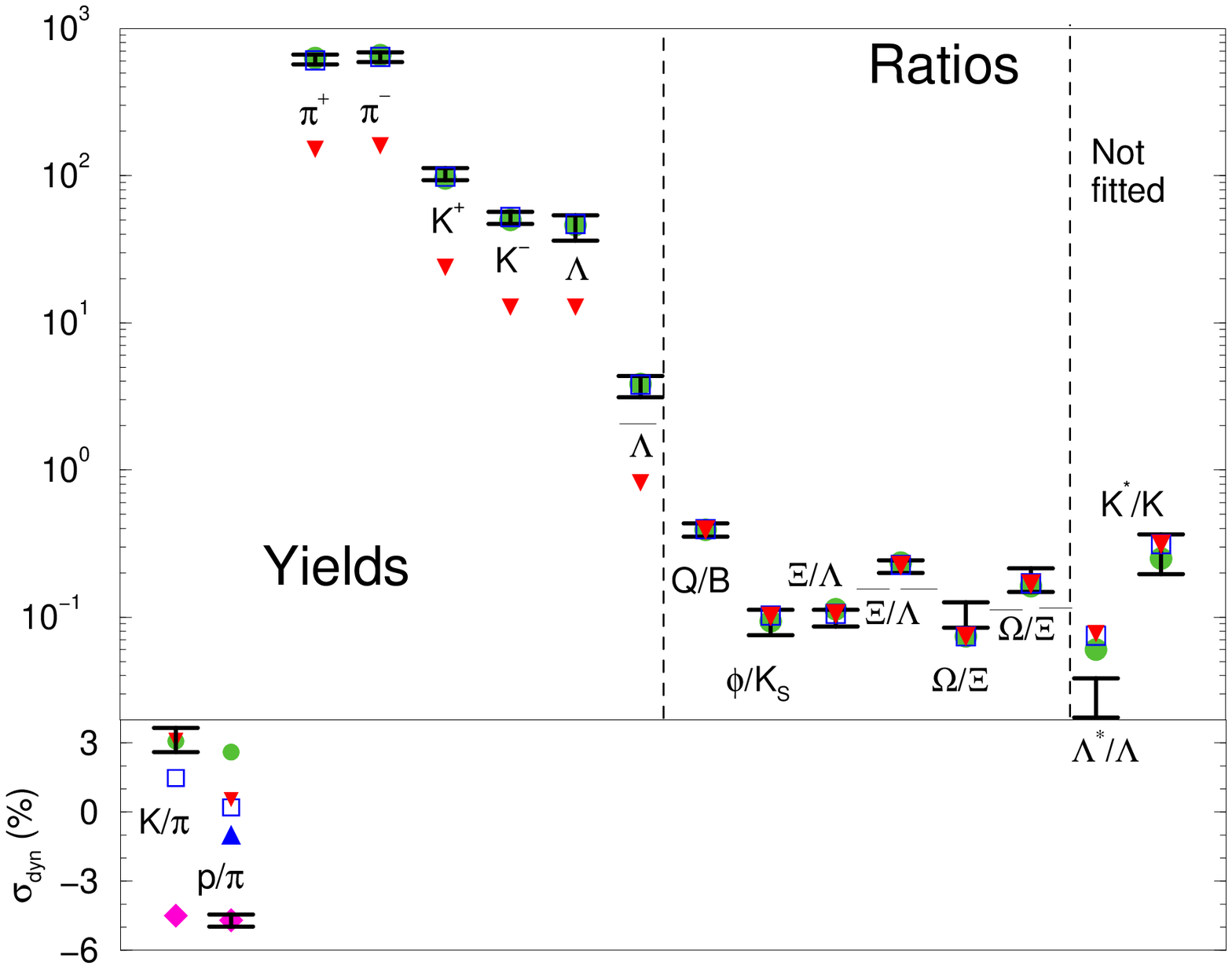}
\caption{\label{graph_200}
(color online) Best fit yields, ratios and fluctuations for RHIC (left panel) and SPS (right panel).  Green circles represent a fit where $\gamma_q$ was fitted.  Red triangles down, $\gamma_q=1$ and yields not fitted.   Blue squares, $\gamma_q=1$ and fluctuation not fitted.   Red triangles, $\gamma_q=1$, normalization determined from the fluctuation and yields not fitted.  The magenta diamond refers to a calculation with halved yields, as per Eq. \ref{caneq}.  The blue triangle up (right panel only) refers to enhanced $\Delta$ production ($m_\Delta=m_p$) }
\end{figure*}

The results of the analysis are shown in table \ref{restable} and Fig. \ref{graph_200}.
An equilibrium analysis ($\gamma_q=1$, empty blue squares in Fig. \ref{graph_200}) fits the particle yields and ratios but under-predicts $\sigma_{K/\pi}^{dyn}$ by many standard deviations.
If $\sigma_{K/\pi}^{dyn}$ is fitted together with ratios (red triangles down in Fig. \ref{graph_200}) but no yields, it forces the system volume (all other parameters being the same within error) to be unrealistically small ($\sim 500 fm^3$ at RHIC, $\sim 1000 fm^3$ at SPS), thereby under-predicting particle yields by several standard deviations.
It is only the addition of $\gamma_q$  (circles in Fig \ref{graph_200}) that allows fluctuations to be driven to a high enough value while
maintaining sufficiently high volume to describe the particle multiplicities, and sufficiently high temperature to describe ratios.

It is important to underline that both yields and fluctuations contribute to such a precise determination:  Equilibrium statistical models can describe
most yields and ratios acceptably with $\gamma_q=\gamma_s=1$, but fail to describe the event-by-event fluctuation.  Conversely, transport models provide an
acceptable description of event-by-event fluctuations \cite{flucttrans}, but fail to describe the yield of multi-strange particles \cite{urqmdstrange}.

Unlike what is sometimes asserted, the $\Lambda(1520)$ and $K^*$ are acceptably described by the statistical model at RHIC.  The strongest disagreement arises
from $\Sigma^*$ under-prediction, at the level of 1.5 standard deviations.   We await for more measurements of resonances such as $\Delta$ and $\rho$ in
central collisions before trying to interpret this under-prediction.  

The acceptable description of the $\Lambda(1520)$ and $K^*$ yield using the same freeze-out temperature as the stable particles, and the {\em under-prediction}
of the $\Sigma^*$, makes a case for the proposition that the re-interaction period between hadronization and freeze-out might be not as significant as generally
thought.   However, the current data is not capable to {\em rule out} such a long-reinteraction period, since the crucial fluctuations, of ratios
correlated by resonances (e.g. $\sigma_{dyn}^{K+/\pi-}$ vs $\sigma_{dyn}^{K+/\pi-}$) are still not available at RHIC and are not precise enough at for such a falsification at SPS.

 It will also be interesting to see if the SPS resonance results (preliminary since 2001 \cite{resosps}, shown in the plot but not included in the fit) are confirmed:  Here, while the $K^*$ abundance is acceptably described by both the equilibrium and the non-equilibrium model, $\Lambda(1520)$ is considerably over-predicted by both through the disagreement with the equilibrium model is larger, due to the higher freeze-out temperature.

The discrepancies between this fit, and earlier fits with $\gamma_q$ \cite{gammaq_energy} are due to
data-set choice.
The lower temperature, and higher $\gamma_q$ at SPS is due to the $\Sigma^*$,$K^*$ and $\Lambda(1520)$ resonance yields at RHIC that push
for a higher temperature, both in equilibrium and non-equilibrium.
Until now, unfortunately, no published resonance result exists in SPS, although $\Lambda(1520)$ and $K^*$ results are available in 
conference proceedings \cite{resosps}.  These are displayed in Fig. \ref{graph_200}, but were not used in the fit.

Qualitatively, the only major difference between SPS and RHIC systems is the higher baryo-chemical potential ($\mu_B$), expected due to the higher initial transparency at the higher energy RHIC collisions.
The introduction of $\gamma_q$ as a fit parameter makes the freeze-out temperature drop to a value
compatible with the QGP super-cooling hypothesis, and decreases the volume by $\sim 30\%$ (a result 
that goes in the right direction to explain the ``HBT puzzle'' \cite{hydroheinz}).     Due to the very different acceptances in the experiments at RHIC and SPS, a direct comparison of the volumes would not be meaningful.

The main disagreement between the data and {\em all} models is the over-prediction of $\sigma^{dyn}_{p/\pi}$ at SPS.   Other works \cite{spsfluct} have suggested that the low value of $\sigma^{dyn}_{p/\pi}$ is indicative of a rich $\Delta$ resonance abundance at chemical freeze-out.    The preliminary observation of an enhanced $\Delta$ production at RHIC \cite{barannikova} ties in well with this picture.
However, this study puts this conjecture in doubt: As can be seen from the figure, even if one assumes that $m_\Delta=m_p$, or $\sim 80\%$ of the photons come from $\Delta$s (blue triangle up in the right panel of Fig. \ref{graph_200}), it is still not enough to account for the discrepancy. 

It is intriguing that {\em halving} the fluctuation term in Eq. \ref{fluctratio}
\begin{equation}
\label{caneq}
\sigma_{N_1/N_2}= \frac{1}{2} \frac{\ave{(\Delta N_1)^2}}{\ave{N_1}^2} 
+ \frac{\ave{(\Delta N_2)^2}}{\ave{N_2}^2} 
- 2 \frac{\ave{\Delta N_1 \Delta N_2}}{\ave{N_1}\ave{ N_2}}. 
\end{equation}
 produces a result compatible with observations (magenta diamond in Fig. \ref{graph_200}) for $\sigma^{dyn}_{p/\pi}$.  This is the scaling expected, in the thermodynamic limit, if the ensemble physically relevant for baryons were canonical rather than GC \cite{nogc1,nogc2}.   The roughness of this estimation, and the preliminary nature of the data-points in question prevents us from drawing conclusions from the result.  We also point out that if this scaling is applied to Kaons (suggesting C ensemble for strangeness) rather than protons
(suggesting C ensemble for Baryon number), the model fails miserably at both SPS and RHIC, as Fig. \ref{graph_200} shows.    We eagerly await more complete and rigorous studies  using the canonical ensemble \cite{nogchg} to further clarify these issues.

In conclusion, we have used preliminary experimental data to show that, at both 200 GeV Au-Au collisions and 17.6 GeV Pb-Pb collisions, the equilibrium model is unable to describe both yields and fluctuations within the same statistical parameters.
The non-equilibrium model, in contrast, succeeds in describing almost all of the yields and fluctuations measured so far at SPS and RHIC, with the parameters expected from a scenario where
non-equilibrium arises through a phase transition from a high entropy state, with super-cooling and oversaturation of phase space.   Some preliminary SPS data-points, such as the $\Lambda(1520)$ yield and the $\sigma^{dyn}_{p/\pi}$, are however significantly over-predicted in both equilibrium and non-equilibrium models.
We await more published data to determine weather the non-equilibrium
model is really capable of accounting for both yields and fluctuations in all light and strange hadrons produced in heavy ion collisions.

Work supported in part by grants from
the Natural Sciences and Engineering research
council of Canada, the Fonds Nature et Technologies of Quebec, and the Tomlinson foundation.  We would like to thank J. Rafelski, C. Gale and S.Jeon for helpful discussions and continued support.
%

%
%
%

\end{document}